%% file: bare_jrnl_new_sample4.tex
\documentclass[lettersize,journal]{IEEEtran}
\usepackage{amsmath,amsfonts}
\usepackage{algorithmic}
\usepackage{algorithm}
\usepackage{array}
\usepackage[caption=false,font=normalsize,labelfont=sf,textfont=sf]{subfig}
\usepackage{textcomp}
\usepackage{stfloats}
\usepackage{url}
\usepackage{verbatim}
\usepackage{graphicx}
\usepackage{cite}
\usepackage{graphicx}
\usepackage{amsmath}
\usepackage{caption}
\usepackage{subcaption}
\usepackage{amssymb}
\usepackage{tabularx}
\usepackage{longtable}
\usepackage[tableposition=top]{caption}
\usepackage[table, svgnames, dvipsnames]{xcolor}
\usepackage{makecell, cellspace, caption}
\usepackage{float}
\usepackage{multirow}
\usepackage{balance}
\usepackage[most]{tcolorbox}
\usepackage{booktabs}
\usepackage{soul}
\usepackage[caption=false,font=normalsize,labelfont=sf,textfont=sf]{subfig}

\begin{document}
\title{Structural and Statistical Audio Texture Knowledge Distillation for Acoustic Classification}

\author{
    Jarin Ritu, Amirmohammad Mohammadi, Davelle Carreiro, Alexandra Van Dine, and Joshua Peeples%
    \thanks{Manuscript received Month XX, 202X; revised Month XX, 202X.}
    \thanks{Jarin Ritu, Amirmohammad Mohammadi, and Joshua Peeples are with the Department of Electrical and Computer Engineering, Texas A\&M University, College Station, TX, USA (e-mail: jarin.ritu@tamu.edu; amir.m@tamu.edu; jpeeples@tamu.edu).}%
    \thanks{Davelle Carreiro and Alexandra Van Dine are with the Massachusetts Institute of Technology Lincoln Laboratory, Lexington, MA, USA (e-mail: davelle.carreiro@ll.mit.edu; alexandra.vandine@ll.mit.edu).}%
    \thanks{DISTRIBUTION STATEMENT A. Approved for public release. Distribution is unlimited. This material is based upon work supported by the Under Secretary of Defense for Research and Engineering under Air Force Contract No. FA8702-15-D-0001 or FA8702-25-D-B002. Any opinions, findings, conclusions or recommendations expressed in this material are those of the author(s) and do not necessarily reflect the views of the Under Secretary of Defense for Research and Engineering.\textsuperscript{\textcopyright} 2025 Massachusetts Institute of Technology. Delivered to the U.S. Government with Unlimited Rights, as defined in DFARS Part 252.227-7013 or 7014 (Feb 2014). Notwithstanding any copyright notice, U.S. Government rights in this work are defined by DFARS 252.227-7013 or DFARS 252.227-7014 as detailed above. Use of this work other than as specifically authorized by the U.S. Government may violate any copyrights that exist in this work.}%
}

% \markboth{IEEE Transactions on Audio, Speech, and Language Processing}%
% {Ritu \MakeLowercase{\textit{et al.}}: Structural and Statistical Audio Texture Knowledge Distillation for Environmental Sound Classification}

% \markboth{Submitted for review }
% {\MakeLowercase{\textit{}}Submitted for review}

% \IEEEpubid{\makebox[\columnwidth]{0000--0000/00\$00.00~\copyright~2024 IEEE \hfill} \hspace{\columnsep}\makebox[\columnwidth]{}}
% \IEEEpubidadjcol

\maketitle

\begin{abstract}
While knowledge distillation has shown success in various audio tasks, its application to environmental sound classification often overlooks essential low-level audio texture features needed to capture local patterns in complex acoustic environments. To address this gap, the Structural and Statistical Audio Texture Knowledge Distillation (SSATKD) framework is proposed, which combines high-level contextual information with low-level structural and statistical audio textures extracted from intermediate layers. To evaluate its generalizability across diverse acoustic domains, SSATKD is tested on four datasets within the environmental sound classification domain, including two passive sonar datasets (DeepShip and Vessel Type Underwater Acoustic Data (VTUAD)) and two general environmental sound datasets (Environmental Sound Classification 50 (ESC-50) and Tampere University of Technology (TUT) Acoustic Scenes). Two teacher adaptation strategies are explored: classifier-head-only adaptation and full fine-tuning. The framework is further evaluated using various convolutional and transformer-based teacher models. Experimental results demonstrate consistent accuracy improvements across all datasets and settings, confirming the effectiveness and robustness of SSATKD in real-world sound classification tasks.\footnote{\thanks{\protect\url{https://github.com/Advanced-Vision-and-Learning-Lab/SSATKD_Lightning}}}
\end{abstract}

\input{Sections/Introduction}
\input{Sections/Related_work}

\input{Sections/Methodology}
\input{Sections/Experimental_result}

\input{Sections/Concluison}
\balance
\bibliographystyle{IEEEtran}
\bibliography{ref}
\end{document}

%% file: Sections/Introduction.tex
\section{Introduction} \label{sect:introduction}
\IEEEPARstart{C}{lassifying} real-world audio signals plays a vital role in various applications, ranging from urban scene analysis to marine monitoring \cite{Xu, wang2014robust}. Environmental sound classification (ESC) solutions help to classify signals relevant to these applications and encompass a wide range of tasks involving acoustic event detection in both terrestrial and underwater settings. Well-studied ESC tasks involve recognizing sounds like vehicle horns, animal vocalizations, or human activities in terrestrial, often cluttered, environments \cite{piczak2015esc}. Another application of ESC involves sonar signal classification, which seeks to identify underwater sound sources such as vessels and marine life by their acoustic signatures \cite{Ghosh, Neupane}. Sonar signal classification is a key construct in marine biology, defense, and underwater infrastructure monitoring applications. 

While there are two modalities of sonar sensing, namely active and passive, this effort focuses on passive sonar, which uses sound waves to assess acoustic signals of interest without active emission of signals. Passive sonar classification presents unique challenges due to the complexities of underwater environments, including low signal-to-noise ratios (SNRs), high variability in acoustic signatures \cite{Ghosh,Neupane}, and signal distortion from propagation conditions \cite{li2024noise}. Despite differences in setting, both terrestrial and underwater environmental sound classification tasks share common challenges such as low SNRs, high variability in acoustic patterns, overlapping sources, and complex temporal structures \cite{Lian, Ghosh, Neupane, li2024noise}. Traditional signal processing techniques, such as low-frequency analyzer and recorder (LOFAR) spectra \cite{luo2023survey} and detection of modulation on noise (DEMON) analysis \cite{hashmi2023novel}, often struggle to distinguish signals of interest in these conditions \cite{Ghosh}. To address this, researchers have increasingly turned to machine learning methods like ensemble learning, where multiple models are combined to improve accuracy \cite{Hinton}.

While effective, ensemble models are often computationally expensive \cite{allen2020towards}, prompting interest in more efficient alternatives such as pruning \cite{gordon2018morphnet}, quantization \cite{wu2016quantized}, and knowledge distillation (KD) \cite{Hinton}. KD compresses deep networks by using soft probabilities (logits) from a large teacher model to guide a smaller student model \cite{Cheng}. These soft labels provide richer supervision than hard labels, helping the student network learn more effectively. This process enables the student to match the teacher’s performance while using fewer resources, making it suitable for deployment on real-time or resource-constrained devices \cite{Cheng, Buciluǎ}. KD has shown success in computer vision \cite{beyer2022knowledge} and language processing (NLP) \cite{huang2024knowledge}, but its use in real-world audio classification, particularly in ESC, remains limited.

Existing KD methods typically emphasize high-level semantic knowledge transfer \cite{Shu, yang2024underwater} but often overlook low-level texture features in intermediate audio representations. These features are essential for capturing local acoustic patterns, such as harmonic structure, noise texture, and modulation characteristics. This limitation is especially pronounced in ESC tasks, where texture-based cues may often play a vital role in fine-grained classification \cite{Salamon2017ESC}. To address this, we propose Structural and Statistical Audio Texture Knowledge Distillation (SSATKD), a novel distillation framework that enables student models to learn not only final predictions but also rich audio texture representations from their teachers. SSATKD integrates two specialized modules: one for capturing structural texture patterns and another for quantifying statistical texture variations. Overall, the contributions of this work are as follows:
\begin{itemize}
\raggedright
    \item In-depth analysis of different knowledge distillation strategies (\textit{e.g.}, teacher-student architectures, loss functions) across four diverse datasets within the environmental sound classification domain
    \item Incorporation of a novel Edge Detection Module for extracting structural texture
    \item Implementation of a characteristic-function-based statistical alignment loss for matching distributions without requiring aligned histogram support
\end{itemize}

%% file: Sections/Related_work.tex
\section{Related Work} \label{sect:related_work}

\subsection{Audio Texture Representation}
In audio data analysis, parallels can be drawn between sound and visual textures \cite{Julesz}, which can be categorized into structural and statistical aspects \cite{Srinivasan}. In 1D audio waveforms, time represents the temporal dimension, and amplitude reflects the intensity of the sound at each time point \cite{LIM}. Temporal dependency in audio signals mirrors the spatial dependency observed in 2D images, wherein neighboring pixels exhibit spatial relationships \cite{LIM}. The repetitive patterns or rhythms in audio signals correspond to structural textures in images, revealing regularity and patterns in the sound. These structural textures capture the arrangement of elements within the signal, providing insights into recurring patterns present in the audio data. Meanwhile, amplitude variations in audio signals reflect statistical textures, similar to the intensity values observed in images \cite{Trevorrow}. These statistical textures encapsulate information about the distribution and variability of signal amplitudes, offering insight into the distribution and characteristics of the audio signal \cite{LIM}.

Recent efforts have explored audio classification algorithms by treating time-frequency representations of audio signals as images, a concept inspired by biologically motivated work on object recognition \cite{Yu,Serre}.  By drawing parallels between the characteristics of images and 2D time-frequency representations, methods such as Gabor filters and wavelet transforms can extract meaningful texture information from the data \cite{Yin, Domingos}. These textures are simpler and more consistent than complex sounds like speech or music, motivating the need for improved sound recognition and auditory representation in acoustic data \cite{McDermott}.

\subsection{Deep Learning Methods}
Recent advancements in acoustic signal classification have leveraged deep learning to automatically extract features from time-frequency representations such as spectrograms, mel-frequency cepstral coefficients (MFCCs), and log-mel spectrograms. These representations are widely used to identify a diverse range of acoustic events of interest. Numerous benchmark datasets, such as Tampere University of Technology (TUT) Acoustic Scenes \cite{heittola2022tau} and Environmental Sound Classification 50 (ESC-50) \cite{piczak2015esc} have driven progress in the field by providing labeled audio clips across different environmental scenarios. Models based on convolutional neural networks (CNNs) have demonstrated strong performance in ESC tasks due to their ability to capture local spectral patterns \cite{tokozume2017learning}, while attention mechanisms and transformer architectures have recently been introduced to improve robustness and contextual understanding in complex acoustic scenes \cite{gong2021ast, tan2023attention}.

Similarly, passive sonar processing applies traditional techniques like LOFAR and DEMON to convert raw underwater acoustic data into spectrograms or cepstral features, which are then fed into neural networks \cite{Lian, Yin}. Deep neural networks (DNNs) have been effective in modeling both spatial and temporal patterns within these time-frequency inputs \cite{doan2020underwater}. Recurrent architectures like long short-term memory (LSTM) networks and hybrid CNN-LSTM combinations further enhance performance by capturing long-term dependencies, which are essential for both temporally evolving sound events in environmental scenes \cite{akter2025hybrid} and sequential ship noise in passive sonar \cite{chen2024ship, hummel2024survey}. However, all of these methods can be computationally expensive and often require large datasets, limiting their application in real-time or resource-constrained environments.

\subsection{Knowledge Distillation vs Transfer Learning}
Transfer learning is a widely used strategy in machine learning, particularly when training data is limited or difficult to collect \cite{Tan, weiss2016survey}. Further, transfer learning has been widely used in audio tasks, such as sound event tagging \cite{Diment}, emotional audio research \cite{Yim, Tan}, and environmental audio event detection using semi-supervised learning \cite{Tripathi}. In these scenarios, pre-trained models from related domains are used to improve performance on new tasks, leveraging the knowledge learned from the source domain. Similarly, ESC methods can suffer mismatches between the source and target domains due to different recording conditions or frequency characteristics that can limit the effectiveness of transfer learning \cite{pons2019training, gong2021psla}. Transfer learning can be even less effective in domains like passive sonar, where acoustic properties differ significantly from general audio data. Underwater signals often exhibit high variability due to factors such as vessel movement, environmental noise, and propagation effects related to salinity, depth, and temperature \cite{liu2021few}.

Models pre-trained on unrelated domains (e.g., speech) may fail to generalize well to sonar signals, leading to suboptimal performance \cite{Tan, Tripathi, cook2013transfer}. Knowledge distillation offers an alternative strategy by compressing task-specific knowledge from a large teacher model into a smaller, efficient student model \cite{choi2022temporal}. Unlike transfer learning, which often retains the original model size and domain, knowledge distillation enables lightweight models to approximate the performance of larger ones on the same task. It has been applied in a wide range of tasks, including computer vision, speech recognition, and increasingly, audio classification \cite{Yim, jung2020knowledge, yang2024underwater}. Knowledge distillation has demonstrated the ability to preserve task-specific performance while significantly reducing computational overhead, making it particularly well-suited for real-time applications in resource-constrained environments.

\begin{figure*}[htb]
    \centering
    \includegraphics[width=0.9\linewidth]{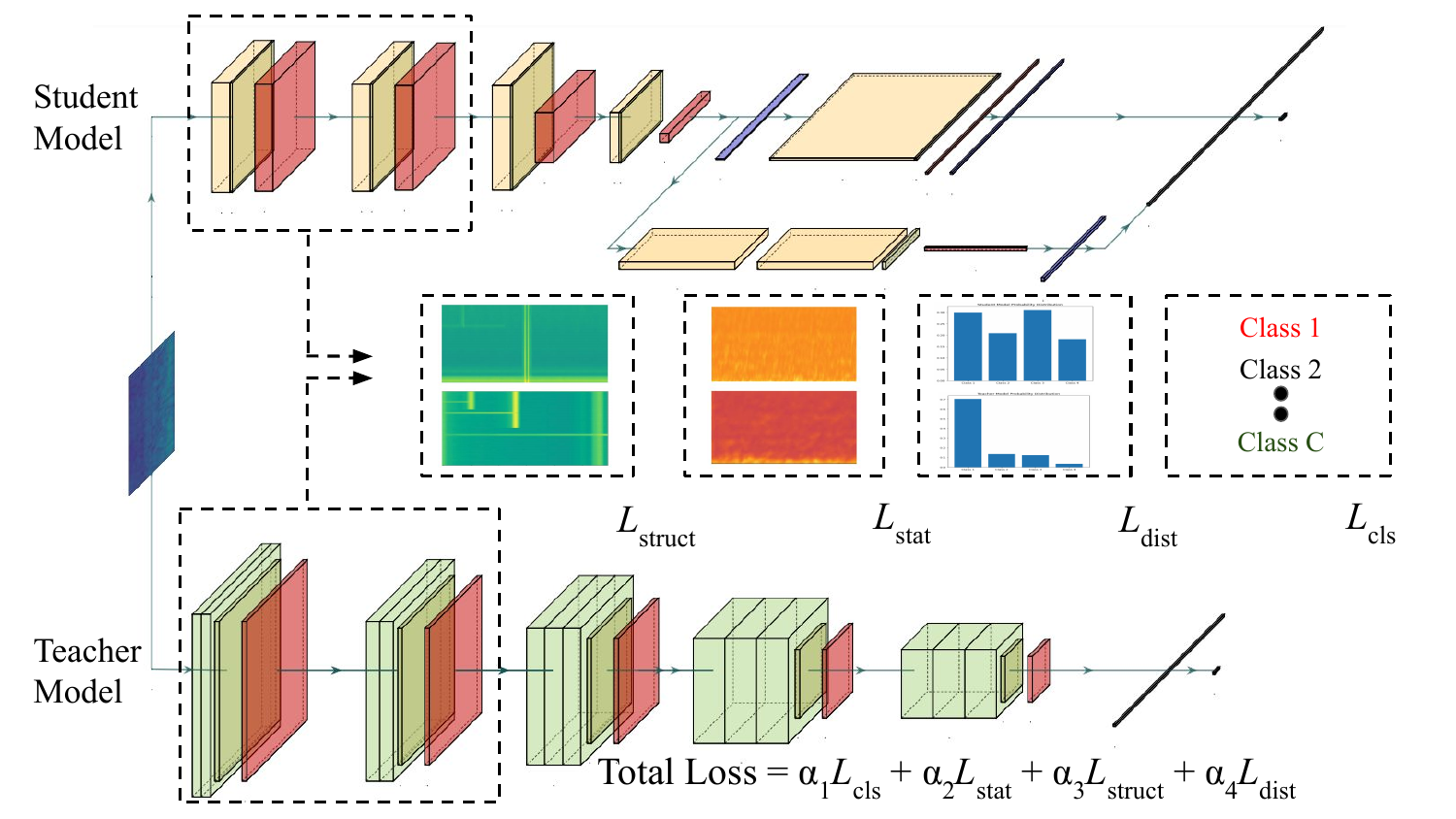}
\caption{Overview of the proposed SSATKD framework. The upper network represents the student model, a Histogram Layer Time Delay Neural Network (HLTDNN) adopted from the original HLTDNN paper \cite{ritu2023histogram}. The framework is designed to be flexible, accommodating any combination of student and teacher models. In this work, the student model is fixed as the HLTDNN, with the teacher model presented as a general structure that can be any of the following pre-trained audio neural network (PANN) model architectures: CNN14, ResNet38, MobileNetV1, or transformer-based foundation models: Wav2Vec 2.0, HuBERT, and Whisper. For CNN-based teachers, each convolutional block includes activation and pooling. Yellow and green blocks represent student and teacher layers, respectively. For transformer-based teachers, features are extracted from early encoder layers and aligned with the student for texture-based distillation. In addition to response-based knowledge distillation, SSATKD incorporates feature-based distillation by extracting texture knowledge from low-level features. Specifically, statistical and structural textures are extracted after the second layer of both the teacher and student models. The total loss is calculated as a weighted sum of classification, statistical, structural, and distillation losses.}

    \label{fig:framework}
\end{figure*}

%% file: Sections/Methodology.tex
\section{Methodology}
The proposed SSATKD framework is illustrated in Figure \ref{fig:framework}. Initially, the input signals are transformed into time-frequency representations, which are then passed into the SSATKD network. The framework focuses on extracting both structural and statistical texture features from the first two layers of the teacher and student models, leveraging the fact that early layers of neural networks capture distinct texture information \cite{ji2022structural}. Following the distillation approach from \cite{Hinton}, the teacher and student networks are aligned by minimizing a combination of response-based and feature-based losses. This alignment ensures the student model effectively learns both high-level semantic information and low-level texture details from the teacher. 
For the statistical texture module, building upon previous research by Zhu et al. \cite{Zhu}, the quantization and count operator (QCO) methodology is refined by replacing their linear binning function with radial basis functions (RBFs) for smoother quantization \cite{ritu2023histogram}. For the structural texture module, a novel Edge Detection Module is introduced, combining hierarchical decomposition techniques, including the Laplacian Pyramid (LP) and edge detection filters.

\subsection{Statistical Texture Module}  
Statistical textures are first extracted by applying Global Average Pooling (GAP) to the input feature matrix \( \mathbf{A} \in \mathbb{R}^{C \times H \times W} \), where \( H \) and \( W \) represent the height and width of the feature map, and \( C \) is the number of channels. This operation results in a global averaged feature vector \( \mathbf{g} \in \mathbb{R}^{C \times 1 \times 1} \), which aggregates information across all spatial dimensions, providing a compact representation of the original feature matrix.  
Next, the cosine similarity between each spatial position \( \mathbf{A}_{ij} \) (where \( i \in [1, W] \) and \( j \in [1, H] \)) in the feature map \( \mathbf{A} \) and the global averaged feature vector \( \mathbf{g} \) is computed, resulting in similarity features \( \mathbf{S} \) with dimensions \( 1 \times H \times W \).  \( \mathbf{S} \) is then reshaped to \( \mathbf{S} \in \mathbb{R}^{H \times W} \) and quantized into \( Q \) levels, denoted as \( \mathbf{Q} = [Q_1, Q_2, \ldots, Q_N] \).
The \( n \)-th quantization level \( Q_n \) is defined in Equation \ref{level}:

\begin{equation}\label{level}
    \mathbf{Q_n} = \frac{n}{N} (\max(\mathbf{S}) - \min(\mathbf{S})) + \min(\mathbf{S}),
\end{equation}
where \( N \) is a hyperparameter for the maximum number of quantization levels, and \( n \in \{1, 2, \ldots, N\} \). 

\begin{figure}[!t]
    \centering
    \includegraphics[width=.45\textwidth]{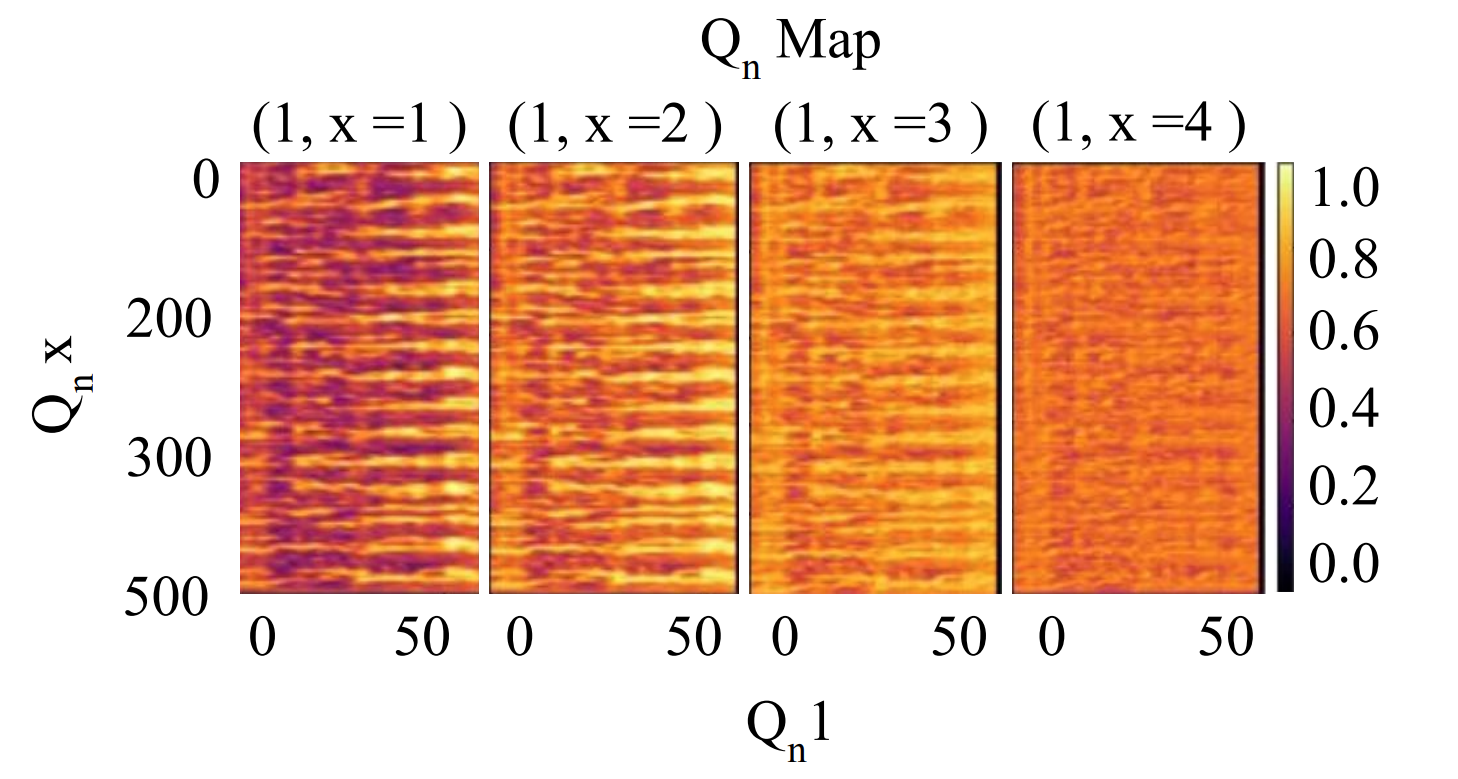}
    \caption{Visualization of 4 co-occurrence matrices out of the 16 possible matrices, corresponding to a 4-level quantization process. Each matrix captures the pairwise quantization co-occurrence between adjacent spectrogram values in the feature maps. The color intensity represents the frequency of co-occurrence for each pair of quantized levels. Brighter regions (yellow to light green) indicate stronger co-occurrences, while darker regions (dark blue and purple) suggest sparse co-occurrences.}
    \label{stat_feature}
\end{figure}

To enhance the encoding process by capturing a smoother gradient compared to the linear basis function used in \cite{Zhu}, the similarity values are further quantized into \( \mathbf{E}_i \in \mathbb{R}^N \) using an RBF. Here, \( i \) ranges from 1 to \( HW \), and each dimension \( n \in \{1, 2, \ldots, N\} \) of \( \mathbf{E}_i \) is calculated using Equation \ref{rbf}. This quantization process is centered around predefined levels and controlled by the bandwidth parameter \( \gamma \), set to \( \frac{1}{N/2} \). 

This choice of \( \gamma \) ensures effective coverage of the interval between the selected centers \( \mathbf{Q} \):

\begin{equation}\label{rbf}
    \mathbf{E}_{i,j} = \exp\left(-{\gamma^2} \left(\mathbf{Q_n} - \mathbf{S}_i \right)^2\right)
\end{equation}

The quantized tensor \( \mathbf{E} \) is then reshaped into \( \mathbb{R}^{N \times 1 \times H \times W} \). For each pair of adjacent spectrogram values in the feature
map,  \( \mathbf{E}_{i,j} \in \mathbb{R}^{N \times 1} \) and \( \mathbf{E}_{i,j+1} \in \mathbb{R}^{N \times 1} \), their outer product \( \hat{\mathbf{E}}_{i,j} \) is computed to capture adjacent information, as defined in Equation \ref{coocurrance}:

\begin{equation}\label{coocurrance}
    \hat{\mathbf{E}}_{i,j} = \mathbf{E}_{i,j} \times \mathbf{E}_{i,j+1}^T
\end{equation}
Here, \( T \) denotes the matrix transpose, and \( \times \) represents matrix multiplication. The resulting co-occurrence matrices for adjacent spectrogram cell pairs are visualized in Figure \ref{stat_feature}, showing how neighboring spectrogram values are correlated. The color scale in the figure ranges from dark purple (representing lower values) to bright yellow (representing higher values). The increasing brightness from the left to right images indicates stronger correlations between adjacent spectrogram cell pairs as moving across the visualizations. This suggests that, in these corresponding regions of \( \mathbf{E} \), the values of neighboring spectrogram values are becoming more similar.

Subsequently, \( \hat{\mathbf{E}} \) is analyzed to generate a 3-D mapping \( \mathbf{C} \in \mathbb{R}^{N \times N \times 3} \), where the first two dimensions represent each possible quantization co-occurrence, and the third dimension signifies the corresponding normalized count. This process is described in Equation \ref{eqn:count}:

\begin{equation}\label{eqn:count}
\mathbf{C} = \text{Concat}\left( \mathbf{Q}, \frac{\sum_{i=1}^{H} \sum_{j=1}^{W} \mathbf{E}_{m,n,i,j}}{\sum_{m=1}^{N} \sum_{n=1}^{N} \sum_{i=1}^{H} \sum_{j=1}^{W} \mathbf{E}_{m,n,i,j}} \right)
\end{equation}
Here, \( \mathbf{Q} \in \mathbb{R}^{N \times N \times 2} \) represents the pairwise combination of all the quantization levels, where \( \mathbf{Q}_{m,n} = [Q_m, Lv_n] \).
The process is summarized in Algorithm \ref{alg:StatisticalTexture}, detailing the steps from cosine similarity calculation to generating the final co-occurrence maps.
\begin{algorithm}
\caption{Statistical Texture Module Processing}
\label{alg:StatisticalTexture}
\begin{algorithmic}
\STATE \textsc{Input:} Feature matrix $ \mathbf{A} \in \mathbb{R}^{C \times H \times W} $, RBF parameter $\gamma$, quantization levels $N$
\STATE \textsc{Output:} 3D co-occurrence map, $ \mathbf{C} \in \mathbb{R}^{N \times N \times 3} $
\STATE

\STATE \textsc{Similarity Estimation}
\STATE \hspace{0.5cm} Compute global averaged vector $ \mathbf{g} \in \mathbb{R}^{C \times 1 \times 1} $ from $\mathbf{A}$
\STATE \hspace{0.5cm} Compute cosine similarity matrix $ \mathbf{S} \in \mathbb{R}^{H \times W} $ between $\mathbf{g}$ and each local vector $ \mathbf{A}_{[i,j]} $
\STATE

\STATE \textsc{Quantization and Embedding}
\STATE \hspace{0.5cm} Quantize $ \mathbf{S} $ into $N$ levels: $ \mathbf{Q} = \{Q_1, \dots, Q_N\} $
\STATE \hspace{0.5cm} Compute RBF-based quantized values $ \mathbf{E} \in \mathbb{R}^{N \times 1 \times H \times W} $
\STATE

\STATE \textsc{Co-occurrence Construction}
\STATE \hspace{0.5cm} Compute outer product $ \hat{\mathbf{E}}_{[i,j]} = \mathbf{E}_{[i,j]} \cdot \mathbf{E}_{[i,j+1]}^{\mathrm{T}} $
\STATE \hspace{0.5cm} Compute co-occurrence map $ \hat{\mathbf{E}} \in \mathbb{R}^{N \times N \times H \times W} $
\STATE \hspace{0.5cm} Aggregate to form 3D map $ \mathbf{C} \in \mathbb{R}^{N \times N \times 3} $
\STATE

\STATE \textbf{return} 3D co-occurrence map, $ \mathbf{C} $
\end{algorithmic}
\end{algorithm}

\subsection{Structural Texture Module}
\begin{figure}[!t]
    \centering
    \includegraphics[width=1\columnwidth]{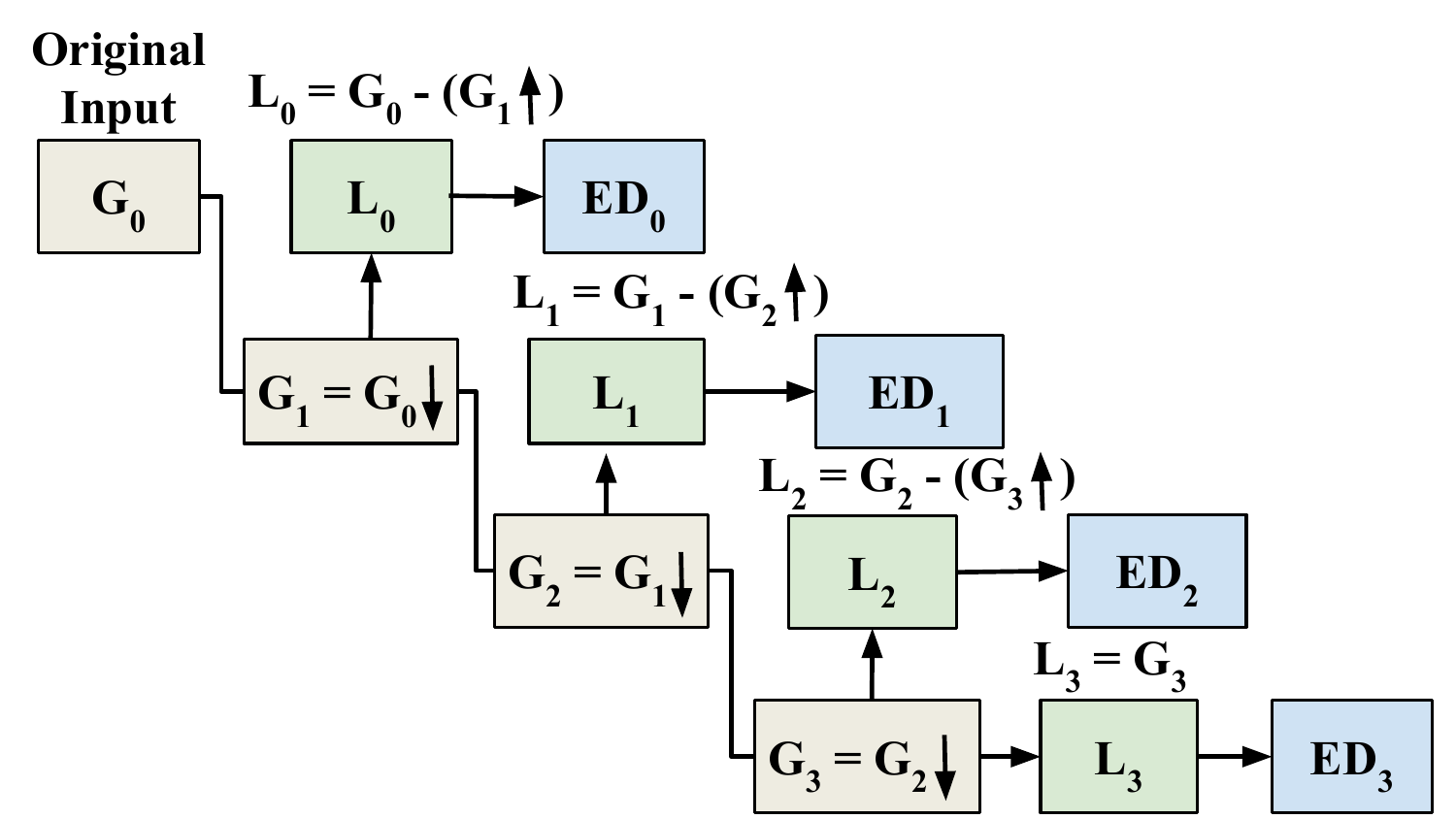}
    \caption{The steps in the structural module. \( \mathbf{L}_0, \mathbf{L}_1, \mathbf{L}_2, \mathbf{L}_3 \) represent the high-pass filtered spectrograms generated by the LP decomposition, while \( \mathbf{G}_0, \mathbf{G}_1, \mathbf{G}_2, \mathbf{G}_3 \) correspond to the low-pass filtered spectrograms produced by the Gaussian Pyramid (GP). \( \mathbf{ED}_0, \mathbf{ED}_1, \mathbf{ED}_2, \mathbf{ED}_3 \) denote the edge detection filters applied at each level.}
    \label{struct_steps}
\end{figure}

% \begin{figure}[!t]
%     \centering
%     \includegraphics[width=1\columnwidth]{Figures/Figure3_struct_steps.pdf}
%     \caption{The steps in the structural module. \( \mathbf{L}_0, \mathbf{L}_1, \mathbf{L}_2, \mathbf{L}_3 \) represent the high-pass filtered spectrograms generated by the LP decomposition, while \( \mathbf{G}_0, \mathbf{G}_1, \mathbf{G}_2, \mathbf{G}_3 \) correspond to the low-pass filtered spectrograms produced by the Gaussian Pyramid (GP). \( \mathbf{ED}_0, \mathbf{ED}_1, \mathbf{ED}_2, \mathbf{ED}_3 \) denote the edge detection filters applied at each level.}
%     \label{struct_steps}
% \end{figure}
Effective texture representation is critical for texture classification, especially when dealing with challenges such as scale variability and complex textural patterns \cite{tang2007extraction}. In the approach used here, structural texture information in the spectral domain is extracted using a novel edge detection module. This module combines hierarchical decomposition techniques, including the Gaussian Pyramid (GP), Laplacian Pyramid (LP), and edge detection filters, as illustrated in Figure \ref{struct_steps}.
\subsubsection{Laplacian Pyramid Decomposition}
\begin{figure}[htb]
    \centering
    \includegraphics[width=\columnwidth]
    {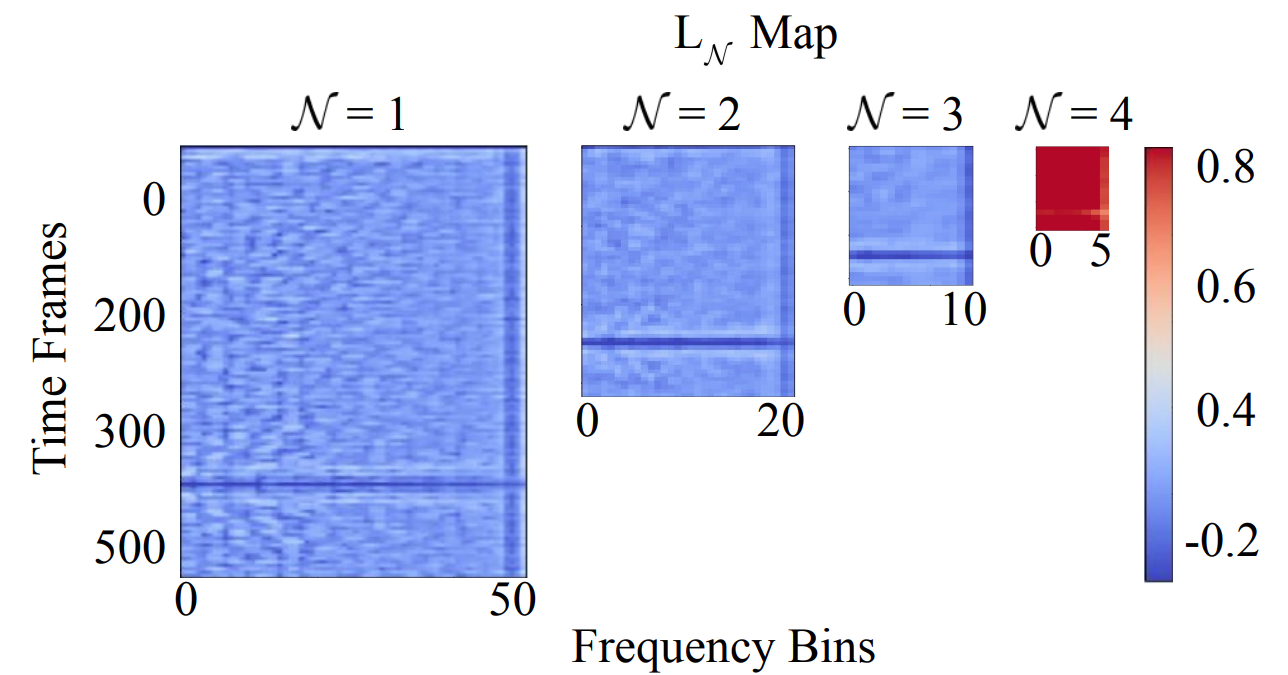}
\caption{Visualization of the 4-level LP decomposition stages. This figure illustrates the downsampling process that generates multi-scale Gaussian levels, with the LP capturing the differences between these levels. The decomposition preserves fine details across different scales, highlighting the transitions from finer to coarser resolutions.}
    \label{fig:decomposition}
\end{figure}

LP is a linear and invertible representation that consists of band-pass images derived from a GP, each representing different scales, along with a low-frequency residual \cite{burt1987laplacian}. The downsampling operation, denoted as \( \downarrow \), blurs and reduces the size of a matrix \( \mathbf{I} \), producing a smaller matrix \( \mathbf{I} \downarrow \) with half the height and width of the original, as shown in Figure \ref{fig:decomposition}. Conversely, the upsampling operation, denoted as \( \uparrow \), smooths and doubles the size of a matrix \( \mathbf{I} \), resulting in a matrix \( \mathbf{I} \uparrow \) with dimensions twice that of the input.
\begin{algorithm}
\caption{Structural Texture Module Processing}
\label{alg:StructuralTexture}
\begin{algorithmic}
\STATE \textsc{Input:} Feature map $\mathbf{I} \in \mathbb{R}^{C \times H \times W}$, levels $\mathcal{N}$, directional Sobel filters
\STATE \textsc{Output:} Structural texture representation, $\mathbf{T} \in \mathbb{R}^{C' \times H \times W}$ \hfill 
\STATE

\STATE \textsc{Laplacian Pyramid Decomposition}
\STATE \hspace{0.5cm} $\mathbf{G}_0 \gets \mathbf{I}$
\FOR{$k = 0$ to $\mathcal{N}-1$}
    \STATE \hspace{0.5cm} $\mathbf{G}_{k+1} \gets \text{Downsample}(\mathbf{G}_k)$
    \STATE \hspace{0.5cm} $\mathbf{L}_k \gets \mathbf{G}_k - \text{Upsample}(\mathbf{G}_{k+1})$
\ENDFOR
\STATE \hspace{0.5cm} $\mathbf{L}_\mathcal{N} \gets \mathbf{G}_\mathcal{N}$ \textit{(low-frequency residual)}
\STATE

\STATE \textsc{Edge Detection}
\FOR{each $\mathbf{L}_k$}
    \FOR{each direction $\theta \in \{0^\circ, 45^\circ, ..., 315^\circ\}$}
        \STATE \hspace{0.5cm} $\mathbf{E}_k^\theta \gets \text{Sobel}(\mathbf{L}_k, \theta)$
    \ENDFOR
\ENDFOR
\STATE

\STATE \textsc{Edge Fusion}
\STATE \hspace{0.5cm} \textit{(Option 1) Weighted Sum:} $\mathbf{T} \gets$ GroupedConv\big($\{\mathbf{E}_k^\theta\}$\big)
\STATE \hspace{0.5cm} \textit{(Option 2) Max Fusion:} $\mathbf{T} \gets \max_{\theta}(\mathbf{E}_k^\theta)$
\STATE \hspace{0.5cm} \textit{(Option 3) All Fusion:} $\mathbf{T} \gets \text{Concat}\big(\mathbf{E}_k^\theta \ \forall k, \theta\big)$
\STATE

\STATE \textbf{return} Structural texture representation, $\mathbf{T}$
\end{algorithmic}
\end{algorithm}

To construct the GP \( \{\mathbf{G}_0, \mathbf{G}_1, \ldots, \mathbf{G}_\mathcal{N}\} \), the downsampling operation is applied iteratively to the original image,\( \mathbf{G}_0 = \mathbf{I} \) iteratively to generate each subsequent level \( \mathbf{G}_k \). In this case, a 4-level decomposition is used, meaning \( \mathcal{N} = 4 \). The LP \( \{\mathbf{L}_0, \mathbf{L}_1, \ldots, \mathbf{L}_{\mathcal{N}-1}\} \) is created by subtracting the upsampled lower-resolution Gaussian level \( \mathbf{G}_{k+1} \uparrow \) from the current level \( \mathbf{G}_k \), as defined in Equation \ref{Laplacian}:

\begin{equation}\label{Laplacian}
    \mathbf{L}_k = \mathbf{G}_k - (\mathbf{G}_{k+1} \uparrow)
\end{equation}

Each level \( \mathbf{L}_k \) captures details at a specific scale, while the final level \( \mathbf{L}_\mathcal{N} \) represents the low-frequency residual, equivalent to the last level of the GP, \( \mathbf{G}_N \). To reconstruct the original image from the LP, the process is reversed, as shown in Equation \ref{Gaussian}:

\begin{equation}\label{Gaussian}
    \mathbf{G}_k = \mathbf{L}_k + (\mathbf{G}_{k+1} \uparrow)
\end{equation}

The reconstruction process begins with \( \mathbf{G}_N = \mathbf{L}_N \) and proceeds by upsampling and adding the difference matrices from finer levels until the full-resolution image \( \mathbf{G}_0 \) is restored. This process is visualized in Figure \ref{fig:decomposition}.

\subsubsection{Edge Detection Filters and Edge Responses}
At each level of the LP, directional information is captured using edge detection filters. Specifically, Sobel kernels \cite{peeples2024histogram} are applied to the feature maps at each level to generate edge responses across eight orientations: 0, 45, 90, 135, 180, 225, 270, and 315 degrees. These edge responses emphasize the directional texture patterns, which are crucial for robust texture representation. To aggregate these edge responses into the final structural texture map $\mathbf{T}$, three distinct fusion methods are implemented:
\begin{itemize}
    \item \textbf{Weighted Sum}: A grouped convolution combines responses, learning their relative importance. 
    \item\textbf{Max}: The strongest edge response is retained by selecting the maximum value across channels. 
    \item\textbf{All}: All edge responses are retained, preserving full directional information.
\end{itemize}
The fused structural texture representation $\mathbf{T} \in \mathbb{R}^{C' \times H \times W}$ captures directional edge patterns across multiple pyramid levels and serves as a rich descriptor for structural texture analysis.

\subsection{Loss Functions in SSATKD}  
 Distinct loss functions are adopted to balance the objectives of texture analysis and classification.  
\subsubsection{Statistical Loss}

The statistical texture module produces a joint co-occurrence 
representation $\mathbf{C} \in \mathbb{R}^{N \times N \times 3}$, where 
the first two dimensions correspond to quantization-level pairs and 
the third dimension stores the normalized co-occurrence count. 
Each bin $(i,j)$ is associated with the coordinate pair 
$\mathbf{x}_{ij} = [Q_i, Q_j]^\top \in \mathbb{R}^2$. 
The resulting joint histogram $\mathbf{H} \in \mathbb{R}^{N \times N}$ 
represents the normalized co-occurrence counts and satisfies

\begin{equation}
\sum_{i=1}^{N} \sum_{j=1}^{N} \mathbf{H}(i,j) = 1,
\label{eq:hist_normalization}
\end{equation}

so that $\mathbf{H}$ defines a discrete probability distribution over the 
2D quantization space. The associated bin-center coordinates are given by 
$\mathbf{x}_{ij} = [Q_i, Q_j]^\top$.

Direct bin-to-bin comparison between teacher and student histograms assumes 
aligned discretization and overlapping support \cite{qin2021structure, avi2023differentiable}. 
In practice, however, small shifts of probability mass across neighboring bins may 
produce large element-wise differences despite representing similar underlying 
distributions. A more principled approach is therefore to compare the 
distributions themselves rather than individual histogram entries. 
Furthermore, in audio analysis, statistical properties of spectrogram-based 
representations are commonly interpreted through frequency-domain principles 
\cite{meriem2024texture}. The characteristic function, defined as the Fourier transform of a probability distribution, provides an alternative representation of the distribution in the characteristic-function (Fourier) domain \cite{feuerverger1977empirical}. 

Rather than comparing histogram bins directly, distributions can be compared through their characteristic functions, which aggregate contributions from all histogram bins and enable a global comparison of the empirical distributions. In this representation, each bin contributes through a complex exponential term whose magnitude is bounded. Consequently, the influence of each bin is limited and varies smoothly with respect to the bin coordinates. As a result, small redistributions of probability mass across neighboring bins produce correspondingly small changes in the representation, reducing sensitivity to discretization effects that can otherwise lead to large element-wise differences in direct histogram comparisons. Similar robustness properties of characteristic-function-based representations have been observed in statistical estimation and learning settings \cite{markatou1995robust}. For these reasons, a characteristic-function-based formulation is adopted.

To approximate multivariate distribution matching efficiently, a set of 
$M$ random unit projection directions 
$\{\mathbf{a}_m\}_{m=1}^{M}$ is sampled, where 
$\mathbf{a}_m \in \mathbb{R}^2$ and $\|\mathbf{a}_m\|_2 = 1$. 
For a projection direction $\mathbf{a}_m$ and frequency $t \in \mathbb{R}$, 
the projected empirical characteristic function of the histogram is defined as

\begin{equation}
\hat{\varphi}(t; \mathbf{a}_m)
=
\sum_{i=1}^{N} \sum_{j=1}^{N}
\mathbf{H}(i,j)
\exp\!\left( \mathrm{i}\, t\, \mathbf{a}_m^\top \mathbf{x}_{ij} \right),
\label{eq:ech}
\end{equation}

where $\mathrm{i}=\sqrt{-1}$ denotes the imaginary unit.

\noindent A fixed set of $K=32$ frequencies $\{t_k\}_{k=1}^{K}$ is uniformly sampled 
from the symmetric interval $[-T_{\max}, T_{\max}]$. 
The statistical loss between teacher and student histograms is defined as

\begin{equation}
L_{\text{stat}}
=
\frac{1}{MK}
\sum_{m=1}^{M}
\sum_{k=1}^{K}
\omega(t_k)
\left|
\hat{\varphi}_T(t_k; \mathbf{a}_m)
-
\hat{\varphi}_S(t_k; \mathbf{a}_m)
\right|^2,
\label{eq: statloss}
\end{equation}

\noindent where $|\cdot|^2$ denotes the squared modulus of a complex number. 
The frequency weighting function is defined as

\begin{equation}
\omega(t_k) = \exp(-\lambda t_k^2),
\label{eq:weight}
\end{equation}

with $\lambda > 0$ controlling attenuation of high-frequency components. 
This Gaussian weighting emphasizes dominant distributional components 
while reducing sensitivity to discretization noise and high-frequency 
fluctuations. The parameters $\lambda$ and $T_{\max}$ are selected 
empirically and kept fixed across all experiments. This projection-based transform-domain formulation enables stable, 
differentiable, and computationally efficient comparison of teacher 
and student statistical representations without requiring explicit 
histogram bin correspondence.
\subsubsection{Structural Loss}  
To quantify the alignment between structural features, the structural loss \( L_{\text{struct}} \) is defined using cosine similarity. Cosine similarity measures the angle between two vectors, focusing on their directional alignment rather than their magnitude, resulting in an effective metric for comparing structural patterns in the feature space.
The objective is to minimize the discrepancy between the structural features of the teacher and student models. Given the feature maps \( \mathbf{F} \in \mathbb{R}^{C \times H \times W} \), the cosine similarity is calculated along the feature dimension. Thus, the structural loss \( L_{\text{struct}} \) is formulated as Equation \ref{eqn:cossim}:
\begin{equation}\label{eqn:cossim}
L_{\text{struct}} = 1 - \text{CosSim}\left(\mathbf{F}^{\text{struct};T}_{i}, \mathbf{F}^{\text{struct};S}_{i}\right)
\end{equation}

Here, \( \mathbf{F}^{\text{struct};T}_{i} \) and \( \mathbf{F}^{\text{struct};S}_{i} \) represent the structural feature maps of the teacher and student models for the \( i \)-th sample, and the cosine similarity is computed along the channel dimension \( C \).

\subsubsection{Classification Loss} 
Cross-entropy loss is used to measure the alignment between predicted probabilities and true class labels, as defined in Equation \ref{eqn:class}:
\begin{equation}\label{eqn:class}
L_{\text{cls}} = - \sum_{i=1}^{C} y_i \log(p_i)
\end{equation}
where \( p \) represents the predicted probability distribution, \( y \) is the one-hot encoded true labels, and \( C \) is the number of classes.

\subsubsection{Distillation Loss} 
To promote response-based knowledge distillation and enable the student to approximate the teacher’s soft targets, Earth Mover’s Distance (EMD) loss is employed \cite{rubner2000earth}. The bin centers and widths are shared between student and teacher, as they represent softmax logits corresponding to class probability distributions. EMD measures the dissimilarity between probability distributions by computing the minimum cost required to transport probability mass from one distribution to another. This formulation enables cross-category comparison of probabilities and captures relationships between classes during distribution alignment \cite{lv2024wasserstein}. The distillation loss, \( L_{\text{distill}} \), is defined in Equation \ref{eqn:distill} as the mean squared difference between the CDFs of the student and teacher models.

% \subsubsection{Distillation Loss} 
% To promote response-based knowledge distillation and enable the student to approximate the teacher’s soft targets, EMD loss is employed. The bin centers and widths are shared between student and teacher, as they represent softmax logits corresponding to class probability distributions. EMD is well-suited for cases where teacher and student outputs differ in shape or smoothness, particularly when direct layer mappings are infeasible due to architectural differences \cite{yeh2021equivariant}. EMD quantifies the dissimilarity between probability distributions by calculating the minimum cost required to transform one distribution into another. The distillation loss, \( L_{\text{distill}} \), is defined in Equation \ref{eqn:distill} as the mean squared difference between the CDFs of the student and teacher models.

\begin{equation}\label{eqn:distill}
    L_{\text{distill}} =   \frac{1}{C} \sum_{i=1}^{C} \left( \text{CDF}_{\text{student}}(i) - \text{CDF}_{\text{teacher}}(i) \right)^2 
\end{equation}
where \(i\) refers to the class index within the CDFs and \(C\) is the number of classes. The sum computes the squared difference between the CDFs of the student and teacher models across \(C\) classes for a single sample.

\subsection{Overall Objective Function}
Achieving the right balance among objectives is crucial for maximizing the model's overall performance. To achieve this balance, an uncertainty-based loss weighting approach \cite{kendall2018multi} is utilized. This method dynamically adjusts the contribution of each loss component based on the uncertainty of the corresponding task. The total loss for the SSATKD model is therefore computed using Equation \ref{eqn:total_loss}:
\begin{equation}\label{eqn:total_loss}
    \text{Total Loss} = \alpha_1 \cdot L_{\text{cls}} + \alpha_2 \cdot L_{\text{stat}} + \alpha_3 \cdot L_{\text{struct}} + \alpha_4 \cdot L_{\text{dist}}
\end{equation}
where \( \alpha_1 \), \( \alpha_2 \), \( \alpha_3 \), and \( \alpha_4 \) are the weights assigned to each loss component. These weights are determined based on the variance of each task's predictions, which reflects the uncertainty in the model's performance for that task. The weights are computed using Equation \ref{eqn:uncertinity}:

\begin{equation}\label{eqn:uncertinity}
    \alpha_i = \frac{1}{\sigma_{L_i}^2} + \log(\sigma_{L_i}^2)
\end{equation}

In this equation, \( \sigma_{L_i}^2 \) represents the variance associated with the \( i \)-th task's loss component, with higher variance indicating greater uncertainty. By scaling \( \alpha_i \) inversely with the precision (the inverse of the variance), the model places more importance on tasks where it has higher confidence and reduces the influence of tasks with greater uncertainty. This uncertainty-based weighting approach eliminates the need for manual tuning of loss weights, allowing the model to automatically balance the different loss components for optimal performance.

%% file: Sections/Experimental_result.tex
\section{Experimental Setup}
\subsection{Data Preparation}
\begin{figure}[H]
    \centering
    \includegraphics[width=\columnwidth]{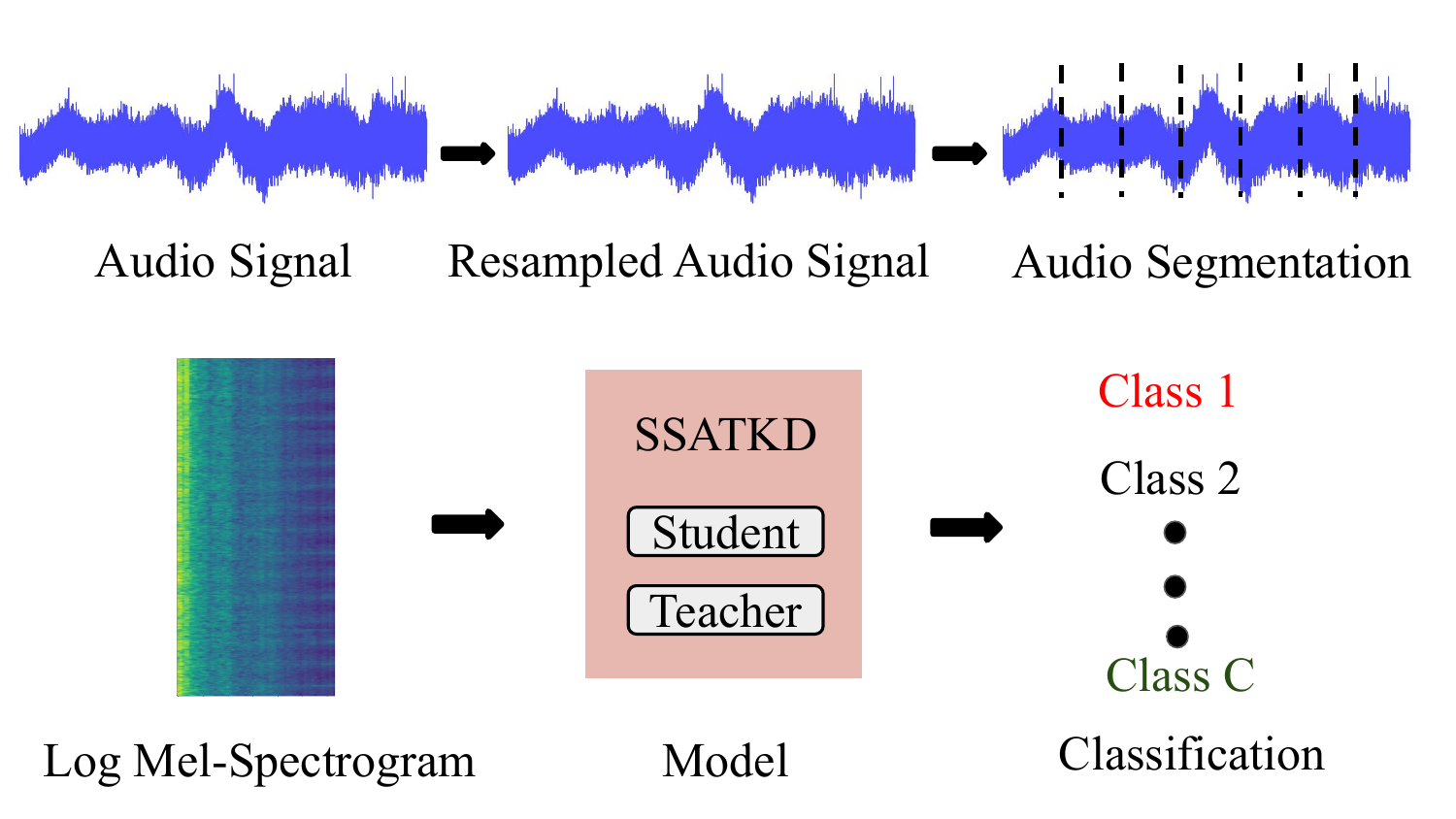}
    \caption{Data preparation pipeline for the SSATKD framework. The process includes resampling audio signals, segmenting them into fixed-length intervals, and converting them into log Mel-frequency spectrograms used as input to the SSATKD framework.}
    \label{fig:dataprep}
\end{figure}

\begin{table}[ht]
    \centering
    \caption{Summary of datasets used in SSATKD experiments.}
    \label{tab:datasets}
    \begin{tabular}{|c|c|c|c|}
        \hline
        Dataset & Total Samples & Classes & Segment Duration \\
        \hline
        DeepShip  & 33,770 & 4 & 5s \\
        \hline
        VTUAD & 175,965 & 5 & 1s \\
        \hline
        ESC-50  & 2,000 & 50 & 5s \\
        \hline
        % UrbanSound8K  & 8,732 & 10 & 4s\\
        % \hline
         TUT Acoustic Scenes & 230,400 & 10 & 1s\\
        \hline
    \end{tabular}
\end{table}

To evaluate the generalizability of the SSATKD framework to a variety of domains, experiments were conducted on four audio classification datasets: two from the underwater acoustic domain (DeepShip \cite{irfan2021deepship} and Vessel Type Underwater Acoustic Data (VTUAD) \cite{nathala2024vessel}) and two from environmental audio (Environmental Sound Classification 50 (ESC-50) \cite{piczak2015esc} and Tampere University of Technology (TUT) Acoustic Scenes 2022 \cite{heittola2022tau}). All audio recordings were resampled to 32 kHz to maintain a consistent sampling rate across datasets. To ensure uniform input dimensions for spectrogram extraction, each recording was either segmented, padded, or truncated to a fixed duration, as summarized in Table \ref{tab:datasets}. 

For the DeepShip dataset, recordings were segmented into non-overlapping 5-second intervals. The dataset was split into training (70\%), validation (20\%), and test (10\%) sets at the recording level, ensuring that segments originating from the same recording do not appear across different splits, thereby preventing data leakage. ESC-50 was used with its predefined 5-fold cross-validation protocol. The TUT Acoustic Scenes dataset was as outlined in the official DCASE 2022 Task 1 setup, employing the predefined splits provided with the dataset. The VTUAD dataset was evaluated in the combined scenario using its predefined training, validation, and test partitions, which consist of 1-second segments. All segments were transformed into log Mel-frequency spectrograms using a Hann window of size 1024, a hop length of 320, and 64 mel filters. SpecAugmentation \cite{park2019specaugment} was applied during training by randomly masking time and frequency bands to improve robustness. These spectrograms served as input to the SSATKD framework. The overall data preparation pipeline is illustrated in Figure~\ref{fig:dataprep}.
\begin{table*}[t]
    \centering
    \caption{Baseline classification accuracy (\%) and model size for the student model (HLTDNN) and teacher models across four datasets. Values represent the average and standard deviation over three independent runs. The best accuracy for each dataset is highlighted in bold.}
    \label{tab:baseline_all}
    \resizebox{\textwidth}{!}{
    \begin{tabular}{|c|c|c|c|c|c|c|}
        \hline
        Model Type & Model Name & \begin{tabular}[c]{@{}c@{}} \# Parameters \end{tabular} & \multicolumn{4}{c|}{Classification Accuracy (\%)} \\
        \cline{4-7}
                   &            &             & DeepShip & VTUAD & ESC-50 & TUT Acoustic Scenes \\
        \hline
        Student Model & HLTDNN       & 11.3K  & 59.62 $\pm$ 1.69 & 80.49 $\pm$ 1.77 & 72.33 $\pm$ 0.76 & 58.13 $\pm$ 1.42 \\
        \hline
        \multirow{6}{*}{Teacher Model}
        & CNN14           & 79.7M  & \textbf{71.33 $\pm$ 1.21} & 96.91 $\pm$ 0.32 & 84.01 $\pm$ 1.21 &  \textbf{67.54 $\pm$ 1.23} \\
        \cline{2-7}
        & ResNet38        & 72.7M  & 64.93 $\pm$ 1.52 & \textbf{97.88 $\pm$ 0.26} & 80.23 $\pm$ 1.33 & 65.34 $\pm$ 1.65 \\
        \cline{2-7}
        & MobileNetV1     & 4.3M   & 66.64 $\pm$ 1.92 & 95.53 $\pm$ 0.33 & 82.07 $\pm$ 1.51 & 66.13 $\pm$ 1.23 \\
        \cline{2-7}
        & Wav2Vec 2.0     & 95.0M  & 63.26 $\pm$ 0.11 & 94.22 $\pm$ 0.84 & 82.31 $\pm$ 1.49 & 65.28 $\pm$ 1.54 \\
        \cline{2-7}
        & HuBERT          & 95.7M  & 63.73 $\pm$ 0.13 & 95.18 $\pm$ 0.71 & 83.56 $\pm$ 1.09 & 65.42 $\pm$ 1.21 \\
        \cline{2-7}
        & Whisper         & 74.0M  & 66.91 $\pm$ 0.18 & 93.35 $\pm$ 0.62 & \textbf{88.84 $\pm$ 1.28} & 65.14 $\pm$ 1.72 \\
        \hline
    \end{tabular}
    }
\end{table*}

\begin{table*}[t]
    \centering
    \caption{SSATKD classification accuracy (\%) with classifier-head-only teacher adaptation. Each teacher model was partially adapted by training only the output layer while freezing the rest of the network. Results are averaged over three independent runs. A 1$\times$1 convolution was added to align teacher and student feature maps, increasing HLTDNN parameters from 11.3K to 12.3K. The last column shows the average gain in accuracy over the baseline HLTDNN across the four datasets.}
    \label{tab:ssatkd_last}
    \resizebox{\textwidth}{!}{
    \begin{tabular}{|c|c|c|c|c|c|c|}
        \hline
        Model & \begin{tabular}[c]{@{}c@{}} \# Parameters \\ (Student vs. Teacher) \end{tabular} & \multicolumn{4}{c|}{Classification Accuracy (\%)} & Avg. Gain (\%) \\
        \cline{3-6}
              &   & DeepShip & VTUAD & ESC-50 & TUT Acoustic Scenes &  \\
        \hline
        Student Only   & 11.3K   & 59.62 $\pm$ 1.69 & 80.49 $\pm$ 1.77 & 72.33 $\pm$ 0.76 & 58.13 $\pm$ 1.42 & N/A \\
        \hline
        Student $+$ CNN14        & 12.3K vs. 79.7M    & 63.58 $\pm$ 1.31 & 83.34 $\pm$ 1.14 & \textbf{77.36 $\pm$ 1.42} &  \textbf{61.41 $\pm$ 1.54} & +3.78 \\
        \hline
        Student $+$ ResNet38     & 12.3K vs. 72.7M    & 63.28 $\pm$ 1.44 & \textbf{85.14 $\pm$ 1.24} & 74.62 $\pm$ 1.32 & 60.16 $\pm$ 1.31 & +3.16 \\
        \hline
        Student $+$ MobileNetV1  & 12.3K vs. 4.3M     & \textbf{65.26 $\pm$ 1.43} & 83.12 $\pm$ 1.22 & 74.58 $\pm$ 1.23 & 61.13 $\pm$ 1.36 & +3.88 \\
        \hline
        Student $+$ Wav2Vec 2.0  & 12.3K vs. 95.0M    & 62.43 $\pm$ 1.31 & 82.44 $\pm$ 1.54 & 75.33 $\pm$ 1.25 & 60.85 $\pm$ 1.55 & +2.62 \\
        \hline
        Student $+$ HuBERT       & 12.3K vs. 95.7M    & 62.75 $\pm$ 1.50 & 83.87 $\pm$ 1.34 & 76.22 $\pm$ 1.43 & 60.77 $\pm$ 1.19 & +3.26 \\
        \hline
        Student $+$ Whisper      & 12.3K vs. 74.0M    & 63.57 $\pm$ 1.16 & 82.82 $\pm$ 1.29 & 75.89 $\pm$ 1.67 & 60.17 $\pm$ 1.57 & +2.97 \\
        \hline
    \end{tabular}
    }
\end{table*}

\subsection{Implementation Details}

The HLTDNN model is fixed as the student model in all experiments. For the teacher model, six architectures were evaluated. Three convolutional models from the Pre-trained Audio Neural
Network (PANNs) family \cite{kong2020panns}: CNN14, ResNet38, and MobileNetV1 were selected due to their strong performance in the original PANN benchmark. In addition, three large-scale audio foundation models were included: Wav2Vec 2.0 (Base) \cite{baevski2020wav2vec}, HuBERT (Base) \cite{hsu2021hubert}, and Whisper (Small) \cite{radford2022whisper}. These models represent transformer-based or hybrid (CNN + transformer) encoder architectures trained on diverse speech or audio tasks, offering an opportunity to study generalization in cross-domain distillation.

To investigate the effect of teacher adaptation, two training regimes were considered: (1) training only the output classification layer of the teacher model while freezing all preceding layers, and (2) full fine-tuning of the teacher model on each target dataset before knowledge distillation. All models are trained across four datasets: DeepShip, VTUAD, ESC-50, and TUT Acoustic Scenes. The AdamW optimizer is used with an initial learning rate of 0.0001, which is adjusted using a learning rate schedule defined in Equation \ref{eqn:lr}:

\begin{equation}
\text{lr} \times \left(1 - \frac{\text{iter}}{\text{max\_iter}}\right)^{0.9}
\label{eqn:lr}
\end{equation}
Training is performed for up to 150 epochs with early stopping based on validation performance, using a patience of 50 epochs and batch size of 32. All experiments are repeated across three independent runs using different random seeds, and average results with standard deviations are reported.

\section{Results and Discussion}

\subsection{Baseline Classification Performance}

Table \mbox{\ref{tab:baseline_all}} summarizes the baseline classification accuracy of the student model, a histogram layer time-delay neural network (HLTDNN), and six teacher models across four datasets. The HLTDNN model, with only 11.3K parameters, consistently achieves reasonable accuracy despite its compact architecture. Its performance is statistically significantly lower than the teacher models, but its low computational costs (\textit{e.g.}, number of learnable parameters) make it suitable for resource-constrained environments. Among the teacher models, CNN14 (79.7M parameters) achieves the highest accuracy on DeepShip (71.33\%) and TUT Acoustic Scenes (67.54\%), demonstrating strong capability in passive sonar and acoustic scene classification tasks. ResNet38 (72.7M parameters) achieves the best performance on VTUAD (97.88\%), indicating its effectiveness in short underwater acoustic signal classification.

On ESC-50, Whisper (74.0M parameters) achieves the highest accuracy (88.84\%), outperforming all other teacher models. This suggests that Whisper’s large-scale pretraining and strong general acoustic representations transfer effectively to environmental sound classification. MobileNetV1, despite having only 4.3M parameters, delivers competitive performance across all datasets, achieving 66.64\% on DeepShip, 95.53\% on VTUAD, 82.07\% on ESC-50, and 66.13\% on TUT. This highlights its favorable balance between efficiency and performance. Notably, CNN14, ResNet38, and MobileNetV1 were pretrained with full supervision on AudioSet \mbox{\cite{gemmeke2017audio}}, which likely contributes to their strong generalization across both environmental and passive sonar datasets.

In contrast, Wav2Vec~2.0 (Base) and HuBERT (Base), which were pretrained using self-supervised learning (SSL), achieve competitive but not top-performing results across the datasets. Although SSL pretraining enables strong general representation learning, these models were not explicitly optimized for audio event classification during pretraining. This may explain why fully supervised AudioSet-pretrained models outperform SSL-based models on certain tasks, particularly passive sonar classification. Overall, the baseline results reveal three key observations: (i) larger fully supervised models such as CNN14 and ResNet38 achieve the strongest performance in passive sonar datasets; (ii) Whisper performs best on ESC-50, demonstrating strong transfer from large-scale speech/audio pretraining; and (iii) the HLTDNN student model offers a highly parameter-efficient alternative at a significantly reduced scale, albeit with lower accuracy. These findings establish a clear performance gap between the compact student model and high-capacity teacher networks, thereby motivating the use of knowledge distillation to transfer rich acoustic representations to the lighter weight HLTDNN architecture.

\subsection{SSATKD Classification Performance}

\subsubsection{Classifier-Head-Only Teacher Adaptation}

Table~\ref{tab:ssatkd_last} presents the performance of the proposed SSATKD framework when each teacher model is adapted using a classifier-head-only strategy. In this setting, only the final output layer of the teacher is fine-tuned on the target dataset, while all earlier layers remain frozen. The student architecture remains HLTDNN across all experiments, with a lighter weight 1$\times$1 convolution added for feature alignment, increasing the parameter count marginally from 11.3K to 12.3K. Even under this constrained adaptation setting, SSATKD consistently improves upon the baseline HLTDNN across all four datasets. On DeepShip, the highest performance is achieved by MobileNetV1 (65.26\%), yielding a substantial improvement over the student-only baseline. For VTUAD, ResNet38 achieves the strongest result (85.14\%), suggesting that deeper convolutional representations remain highly effective for short-duration underwater acoustic signals, even when only the classifier head is adapted.

On ESC-50, CNN14 provides the best performance (77.36\%), indicating that PANN-based architectures pretrained with full supervision retain strong general environmental sound representations under partial adaptation. For TUT Acoustic Scenes, CNN14 again achieves the highest accuracy (61.41\%), demonstrating robustness in acoustic scene classification. Across teachers, the average gain over the baseline ranges from +2.62\% (Wav2Vec~2.0) to +3.78\% (CNN14). Notably, convolutional teachers pretrained with full supervision on AudioSet generally yield larger improvements compared to self-supervised transformer-based models such as Wav2Vec~2.0 and HuBERT. Nevertheless, all teachers provide consistent performance gains, confirming that SSATKD effectively transfers useful acoustic representations even when teacher adaptation is limited. Overall, these results demonstrate that SSATKD remains effective under minimal teacher adaptation. The ability to enhance a student model without fully fine-tuning large teacher networks makes this approach computationally efficient and practical for deployment scenarios with restricted training resources.

\begin{table*}[t]
    \centering
    \caption{SSATKD classification accuracy (\%) with full teacher fine-tuning. Each teacher model was fully adapted to the target dataset prior to distillation. Results are averaged over three independent runs. The last column shows the average gain in accuracy over the baseline HLTDNN across the four datasets.}
    \label{tab:ssatkd_full}
    \resizebox{\textwidth}{!}{
    \begin{tabular}{|c|c|c|c|c|c|c|}
        \hline
         Model & \begin{tabular}[c]{@{}c@{}} \# Parameters \\ (Student vs. Teacher) \end{tabular} & \multicolumn{4}{c|}{Classification Accuracy (\%)} & Avg. Gain (\%) \\
        \cline{3-6}
                      &                                   & DeepShip & VTUAD & ESC-50 & TUT Acoustic Scenes &  \\
        \hline
        Student Only   & 11.3K   & 59.62 $\pm$ 1.69 & 80.49 $\pm$ 1.77 & 72.33 $\pm$ 0.76 & 58.13 $\pm$ 1.42 & N/A \\
        \hline
        Student $+$ CNN14          & 12.3K vs. 79.7M    & 65.76 $\pm$ 1.40 & 86.55 $\pm$ 0.83 & \textbf{82.65 $\pm$ 0.71} & \textbf{63.54 $\pm$ 1.23} & +6.98 \\
        \hline
        Student $+$ ResNet38       & 12.3K vs. 72.7M    & 65.72 $\pm$ 1.48 & \textbf{86.87 $\pm$ 0.74} & 81.71 $\pm$ 0.57 & 62.34 $\pm$ 1.65 & +6.52 \\
        \hline
        Student $+$ MobileNetV1    & 12.3K vs. 4.3M     & \textbf{67.48 $\pm$ 0.86} & 85.37 $\pm$ 1.04 & 81.76 $\pm$ 0.36 & 63.13 $\pm$ 1.23 & +6.79 \\
        \hline
        Student $+$ Wav2Vec 2.0    & 12.3K vs. 95.0M    & 64.87 $\pm$ 1.48 & 85.17 $\pm$ 0.65 & 79.26 $\pm$ 1.05 & 62.28 $\pm$ 1.54 & +5.25 \\
        \hline
        Student $+$ HuBERT         & 12.3K vs. 95.7M    & 66.37 $\pm$ 1.05 & 86.54 $\pm$ 0.62 & 79.26 $\pm$ 1.05 & 62.42 $\pm$ 1.21 & +6.00 \\
        \hline
        Student $+$ Whisper        & 12.3K vs. 74.0M    & 66.78 $\pm$ 1.23 & 85.24 $\pm$ 0.72 & 80.21 $\pm$ 1.26 & 61.14 $\pm$ 1.72 & +5.70 \\
        \hline
    \end{tabular}
    }
\end{table*}
\begin{figure*}
    \centering
    \subfloat[\scriptsize Baseline HLTDNN (59.62 $\pm$ 1.69\%)]{
        \includegraphics[width=0.30\textwidth]{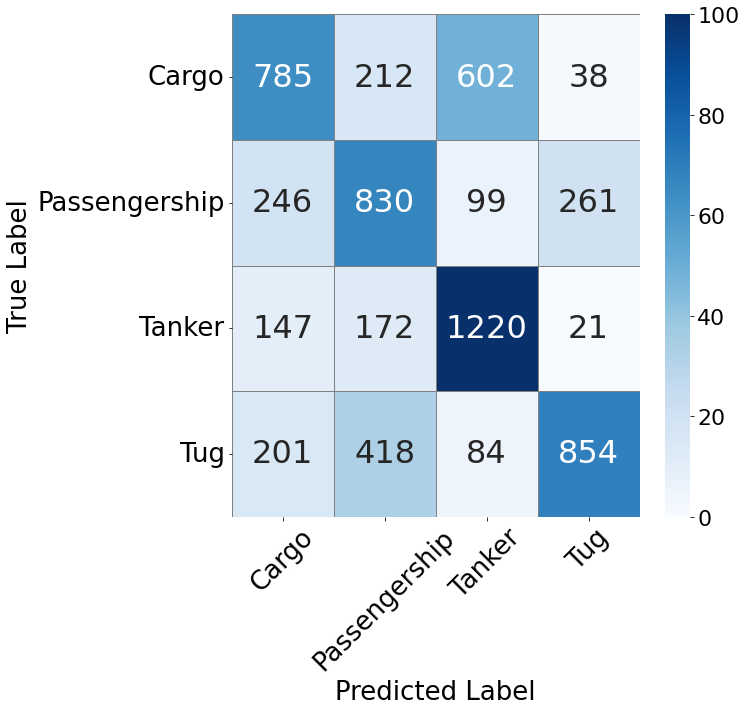}
    }
    \hfill
    \subfloat[\scriptsize SSATKD Classifier-Only (65.26 $\pm$ 1.43\%)]{
        \includegraphics[width=0.30\textwidth]{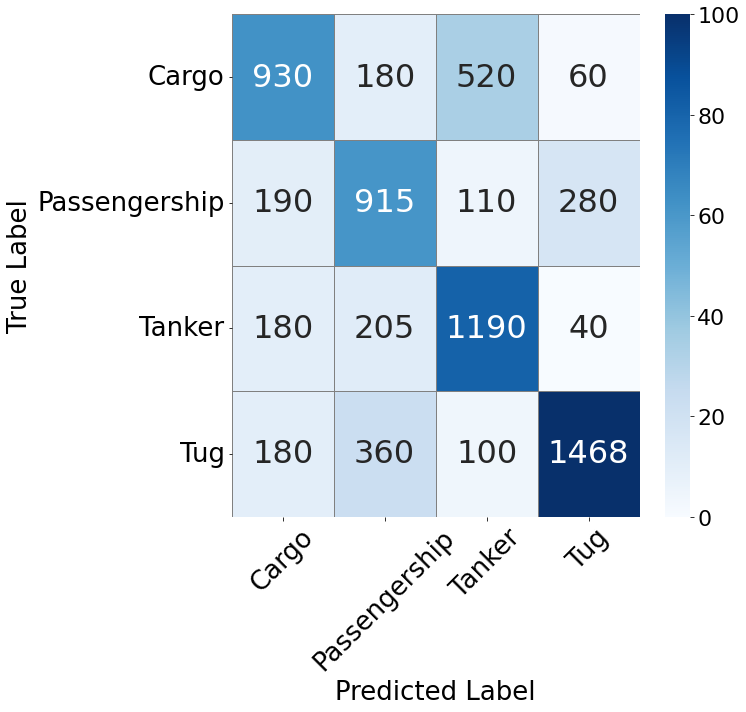}
    }
    \hfill
    \subfloat[\scriptsize SSATKD Full Fine-Tuning (67.48 $\pm$ 0.86\%)]{
        \includegraphics[width=0.30\textwidth]{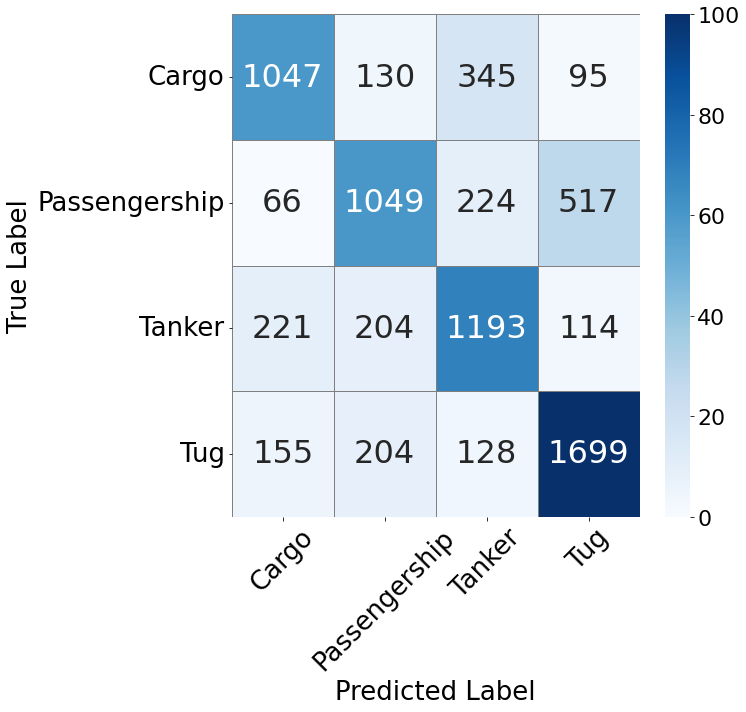}
    }
    \caption{Performance comparison of HLTDNN baseline and SSATKD under two adaptation settings on DeepShip. (a) Baseline HLTDNN. (b) Classifier-only adaptation. (c) Full fine-tuning.}
    \label{fig:ssatkd_comparison}
\end{figure*}

\subsubsection{Full Teacher Fine-Tuning}

Table~\ref{tab:ssatkd_full} presents the classification results when each teacher model is fully fine-tuned on the target dataset prior to applying SSATKD. Compared to the classifier-head-only setting, full teacher adaptation consistently yields larger improvements across all datasets, demonstrating the benefit of transferring task-specialized representations to the lighter weight student model. On the DeepShip dataset, MobileNetV1 achieves the highest student accuracy (67.48\%), followed closely by Whisper (66.78\%) and HuBERT (66.37\%). These results indicate that both compact convolutional architectures and large transformer-based models can effectively transfer domain-adapted knowledge when fully fine-tuned. For VTUAD, ResNet38 achieves the best performance (86.87\%), marginally outperforming CNN14 and HuBERT. This suggests that deep residual convolutional architectures remain highly effective for short underwater acoustic signal classification when fully adapted to the target domain.

In environmental sound classification, CNN14 achieves the highest accuracy on ESC-50 (82.65\%), highlighting the strength of fully supervised AudioSet-pretrained PANN architectures in general acoustic event recognition. On TUT Acoustic Scenes, CNN14 again attains the top performance (63.54\%), demonstrating robustness in acoustic scene modeling. Across teachers, the average gain over the HLTDNN baseline ranges from +5.25\% (Wav2Vec~2.0) to +6.98\% (CNN14), substantially higher than the gains observed under classifier-head-only adaptation. These results confirm that full fine-tuning enables richer domain-specific feature alignment between teacher and student representations, thereby enhancing the effectiveness of SSATKD.

\medskip
\noindent
\textbf{Comparison Between Adaptation Strategies.}
The evaluation of classifier-only and full fine-tuning settings highlights the flexibility of the SSATKD framework under different computational budgets. On the DeepShip dataset with MobileNetV1 as the teacher, the HLTDNN baseline achieves $59.62\% \pm 1.69\%$. Applying SSATKD with classifier-only adaptation improves performance to $65.26\% \pm 1.43\%$, corresponding to a gain of $+5.64$ percentage points. Full fine-tuning further increases accuracy to $67.48\% \pm 0.86\%$, yielding a total improvement of $+7.86$ points over the baseline. Notably, full fine-tuning also reduces performance variance, indicating improved training stability and better feature alignment. While classifier-only adaptation effectively refines the decision boundary with minimal computational cost, full fine-tuning enables deeper feature adaptation, leading to superior overall performance.

\medskip
\noindent\textbf{Class-Wise Performance Analysis.}
To analyze class-level behavior, Fig.~\ref{fig:ssatkd_comparison} presents the confusion matrices for the DeepShip dataset using MobileNetV1 as the teacher: (a) baseline HLTDNN, (b) SSATKD with classifier-only adaptation, and (c) SSATKD with full fine-tuning.
The baseline model shows noticeable confusion between Cargo and Tanker, as well as between Passengership and Tug. In particular, Cargo samples are frequently misclassified as Tanker, and Passengership exhibits substantial confusion toward Tug. These patterns indicate overlapping acoustic characteristics among these vessel types. With classifier-only adaptation, the overall distribution becomes more balanced. While some diagonal entries decrease compared to the baseline, several off-diagonal errors, particularly those with significant misclassification, are reduced. However, inter-class confusion between acoustically similar categories remains evident. Full fine-tuning produces the clearest class separation. Tanker recognition improves further (increased diagonal dominance), and misclassification as unrelated classes is significantly reduced. Although some confusion between Cargo and Tanker and between Passengership and Tug persists, the overall error distribution becomes more structured and less scattered. This improved diagonal concentration directly contributes to the higher overall accuracy reported in Table~\ref{tab:ssatkd_full}.

\subsection{Impact of Loss Components}

To analyze the contributions of individual SSATKD components, an ablation study was conducted across the four datasets covering both passive sonar and environmental sound classification tasks. Analysis examines the impact of the distillation objective, the interaction between structural and statistical texture modules, and key architectural design choices.

\begin{table}[ht]
    \centering
    \caption{Effect of statistical, structural, and distillation loss combinations on student model accuracy across four datasets. Results are reported as classification accuracy (\%) averaged over six teacher–student distillation pairs and three independent runs. The highest accuracy per dataset is shown in bold.
}
    \label{tab:loss}
    \resizebox{\columnwidth}{!}{
    \begin{tabular}{|c|c|c|c|c|c|c|}
         \hline
         Statistical & Structural & Distillation 
         & DeepShip & VTUAD & ESC-50 & TUT \\
         \hline
         &  &  & 59.62 $\pm$ 1.69 & 80.49 $\pm$ 1.77 & 72.33 $\pm$ 0.76 & 58.13 $\pm$ 1.42 \\
         \hline
         &  & \checkmark & 63.62 $\pm$ 1.55 & 81.96 $\pm$ 1.59 & 76.86 $\pm$ 1.74 & 59.21 $\pm$ 1.87 \\
         \hline
         & \checkmark &  & 64.65 $\pm$ 1.49 & 83.11 $\pm$ 1.57 & 77.98 $\pm$ 1.43 & 60.54 $\pm$ 1.42  \\
         \hline
         & \checkmark & \checkmark & 64.91 $\pm$ 1.44 & 83.62 $\pm$ 1.44 & 78.36 $\pm$ 1.55  & 61.26 $\pm$ 1.45  \\
         \hline
         \checkmark &  &  & 65.12 $\pm$ 1.13 & 84.34 $\pm$ 1.30 & 78.58 $\pm$ 1.27 & 61.45 $\pm$ 1.42  \\
         \hline
         \checkmark &  & \checkmark & 65.45 $\pm$ 1.24 & 84.84 $\pm$ 0.99 & 79.19 $\pm$ 1.28 & 61.87 $\pm$ 1.17  \\
         \hline
         \checkmark & \checkmark &  & 65.70 $\pm$ 1.57 & 85.39 $\pm$ 1.27 & 79.77 $\pm$ 1.17 & 62.29 $\pm$ 1.68 \\
         \hline
         \checkmark & \checkmark & \checkmark & \textbf{66.16 $\pm$ 1.18} & \textbf{85.96 $\pm$ 0.77} & \textbf{80.94 $\pm$ 0.84} & \textbf{62.51 $\pm$ 1.43} \\
         \hline
    \end{tabular}
    }
\end{table}

Table~\ref{tab:loss} presents the classification accuracy (\%) across four datasets, with results averaged over six teacher-student distillation pairs and three independent runs per pair. The averaging ensures that the trends discussed below are stable and not due to a specific teacher choice or random initialization. Each individual loss term improves performance over the baseline configuration. When only the statistical loss is added, the model achieves gains of +5.50\% on DeepShip, +3.85\% on VTUAD, +6.25\% on ESC-50, and +3.56\% on TUT. Consistent improvements across all datasets indicate that aligning feature distributions between teacher and student provides useful supervisory information beyond standard classification training. By encouraging the student to match the overall distribution of teacher activations, the model captures richer feature characteristics that are not directly enforced by label supervision alone. Similarly, using only the structural loss also leads to clear improvements: +5.03\% on DeepShip, +2.62\% on VTUAD, +6.28\% on ESC-50, and +3.04\% on TUT. 

The magnitude of these gains is comparable to that of the statistical loss, showing that preserving structural information inside feature maps is equally important. When the distillation loss is applied alone, performance also improves compared to the baseline, although the gains are generally smaller than those obtained with statistical or structural losses. This behavior is reasonable because distillation primarily aligns the final output probabilities of the teacher and student. While this provides useful guidance at the decision level, it does not directly constrain intermediate feature representations. Combining two losses generally produces further improvements over using a single loss. In particular, the joint use of statistical and structural losses leads to stronger and more stable gains across datasets. 

This suggests that the two losses provide complementary information. Since they target different aspects of representation learning, their combination helps the student learn a more complete approximation of the teacher. Adding the distillation loss on top of intermediate alignment (statistical and/or structural) further improves performance. This indicates that output-level guidance and feature-level alignment serve different roles. Feature-level losses shape how internal representations are formed, while distillation ensures that the final predictions remain consistent with the teacher. Finally, the best results on all four datasets are achieved when statistical, structural, and distillation losses are used together. The full combination reaches 66.16\% on DeepShip, 85.96\% on VTUAD, 80.94\% on ESC-50, and 62.51\% on TUT. These correspond to absolute improvements of +6.49\%, +5.47\%, +8.61\%, and +4.38\% over the baselines, respectively. Each loss component contributes useful information, and their integration leads to the most reliable and consistent improvements across diverse acoustic domains.

\subsection{Distillation Loss Comparison}
To investigate the influence of the distillation objective, we compare the commonly used Kullback–Leibler divergence (KLDiv) loss with the Earth Mover’s Distance (EMD) loss. Both losses aim to align the output probability distributions of the teacher and student models; however, they differ fundamentally in how distributional discrepancies are measured. KLDiv penalizes point-wise differences between probability values, whereas EMD considers the underlying geometry of the distribution space by measuring the minimal cost required to transform one distribution into another.

\begin{table}[ht]
    \centering
    \caption{Comparison of KLDiv and EMD as distillation loss functions across four datasets. Results are reported as classification accuracy (\%) of the SSATKD model distilled from six teachers. The better result per dataset is shown in bold.}
    \label{tab:distillation_loss}

    \resizebox{\columnwidth}{!}{
    \begin{tabular}{|c|c|c|c|c|}
         \hline
         Distillation Loss & DeepShip & VTUAD & ESC-50 & TUT \\
         \hline
         KLDiv 
         & 62.65 $\pm$ 1.65
         & 80.77 $\pm$ 1.71
         & 73.83 $\pm$ 1.62
         & 59.07 $\pm$ 1.69 \\
         \hline
         EMD 
         & \textbf{63.73 $\pm$ 1.34}
         & \textbf{81.88 $\pm$ 1.45}
         & \textbf{75.44 $\pm$ 1.58}
         & \textbf{59.53 $\pm$ 1.23} \\
         \hline
    \end{tabular}
    }
\end{table}
\mbox{Table~\ref{tab:distillation_loss}} reports the classification accuracy averaged across six teacher models for each dataset. EMD achieves higher mean accuracy than KLDiv across all four datasets, with improvements of +1.08\% on DeepShip, +1.11\% on VTUAD, +1.61\% on ESC-50, and +0.46\% on TUT. In addition, EMD exhibits comparable or slightly lower variance across most datasets, suggesting more stable optimization behavior. Unlike KLDiv, which measures point-wise divergence between probability values, EMD accounts for the geometric structure of the output distribution by modeling the cost of redistributing probability mass. This property may be particularly beneficial in acoustic classification, where semantically related classes often exhibit correlated prediction distributions. Overall, although the performance gains are moderate, the consistent improvements and the geometry-aware formulation of EMD support its use as the distillation objective within the SSATKD framework.

\subsection{Comparison with Existing State-of-the-Art Knowledge Distillation Methods}
To assess the effectiveness of SSATKD, we compare it with five representative state-of-the-art knowledge distillation (KD) methods: Semantic Representational Distillation (SRD) \cite{yang2024knowledge}, Teacher-Free Knowledge
Distillation (TF-KD) \cite{yuan2020revisiting}, Contrastive Representation Distillation (CRD) \cite{tiancontrastive}, Prime-Aware Adaptive Distillation (PAD) \cite{zhang2020prime}, and Correlation Congruence for Knowledge Distillation
(CCKD) \cite{peng2019correlation}. All methods were implemented using the \texttt{torchdistill} framework \cite{matsubara2021torchdistill} under identical training settings to ensure a fair and consistent evaluation protocol. 

\begin{table}[ht]
    \centering
\caption{Performance comparison of SSATKD with various knowledge distillation methods using the \texttt{torchdistill} framework \cite{matsubara2021torchdistill} across four datasets. Reported results correspond to classification accuracy (\%) of the student model after distillation, averaged over six teacher architectures and three independent runs. The best performance for each dataset is highlighted in bold.}

    \label{tab:KD_com}
    \resizebox{\columnwidth}{!}{
    \begin{tabular}{|c|c|c|c|c|}
        \hline
        Method & DeepShip & VTUAD & ESC-50 & TUT\\ \hline
        SRD \cite{yang2024knowledge}   & 61.15 $\pm$ 1.41\% & 82.41 $\pm$ 1.14\% & 76.22 $\pm$ 1.12\% & 58.97 $\pm$ 1.83\% \\ \hline
        TF-KD \cite{yuan2020revisiting} & 60.13 $\pm$ 1.13\% & 81.51 $\pm$ 1.10\% & 75.46 $\pm$ 1.37\% & 59.12 $\pm$ 1.32\% \\ \hline
        CRD \cite{tiancontrastive} & 61.41 $\pm$ 1.10\% & 81.38 $\pm$ 1.13\% & 76.81 $\pm$ 1.33\% & 60.19 $\pm$ 1.07\% \\ \hline
        PAD \cite{zhang2020prime}   & 57.34 $\pm$ 1.32\% & 78.75 $\pm$ 1.52\% & 72.77 $\pm$ 1.21\% & 56.13 $\pm$ 1.28\% \\ \hline
        CCKD \cite{peng2019correlation}  & 62.18 $\pm$ 1.45\% & 82.49 $\pm$ 1.52\% & 76.25 $\pm$ 1.42\% & 59.53 $\pm$ 1.13\% \\ \hline
        SSATKD (Ours) & \textbf{65.49 $\pm$ 1.15\%} & \textbf{85.49 $\pm$ 1.07\%} & \textbf{81.17 $\pm$ 0.83\%} & \textbf{62.18 $\pm$ 1.43\%} \\ \hline
    \end{tabular}
    }
\end{table}

\noindent
Table~\ref{tab:KD_com} presents a comprehensive comparison between SSATKD and five representative knowledge distillation methods across all benchmarks, with the most pronounced gains observed on ESC-50 (+4.36\%) and DeepShip (+3.31\%). These results highlight important limitations of existing distillation paradigms when applied to complex acoustic signals.

Methods such as CRD and CCKD primarily emphasize global relational structures or second-order feature correlations. While effective in image-based tasks, such approaches may be less expressive in acoustic domains like DeepShip, where different vessel types produce overlapping spectral signatures and subtle temporal variations. In these scenarios, preserving only relational consistency may not fully capture fine-grained structural textures within intermediate feature representations. By explicitly modeling multi-scale spatial and channel dependencies, SSATKD captures nuanced rhythmic and texture-based structural patterns that are critical for distinguishing acoustically similar classes.

TF-KD focuses on softening the teacher’s output distribution, and PAD emphasizes sample weighting strategies. However, in high-variability datasets such as ESC-50, discriminative information is often embedded within intermediate feature distributions rather than solely in final class probabilities. Output-level alignment alone may therefore limit the transfer of intra-class variability. The statistical module in SSATKD addresses this by aligning feature activation distributions, helping preserve representational diversity and preventing excessive compression of the student’s latent space.

Even in datasets with more structured acoustic patterns, such as VTUAD, SSATKD improves upon the strongest baseline by +3.00\%. This suggests that traditional feature alignment captures general signal characteristics but may not retain the full statistical richness of the teacher’s representations. SSATKD enables the student to inherit not only classification decisions but also the structural and statistical properties underlying those decisions. Overall, results indicate that effective distillation for acoustic classification benefits from jointly modeling structural texture and statistical diversity in addition to output alignment. This integrated supervision allows the student network to acquire richer and more expressive representations, leading to consistent improvements across diverse acoustic domains.

%% file: Sections/Concluison.tex
% \section{Conclusion}
% This work introduced SSATKD, a texture-aware knowledge distillation framework that transfers both low-level audio textures and high-level semantic responses from teacher to student models. The results highlight that combining structural and distillation losses yields the highest gains, with response-based distillation playing a dominant role in performance. Comparisons with recent knowledge distillation methods further validated the effectiveness of SSATKD’s hybrid texture-response transfer strategy. This framework has potential to be beneficial for real-time environmental monitoring systems or edge devices where large models are impractical. Future work could explore adaptive loss balancing, expand to other domains like bioacoustics, and integrate self-supervised or multimodal learning to improve generalization and deployment efficiency. 

\section{Conclusion}

This work introduced SSATKD, a texture-aware knowledge distillation framework designed to transfer both low-level audio texture information and high-level semantic responses from teacher to student models for acoustic classification tasks. Unlike conventional distillation approaches that primarily focus on response-level knowledge transfer, SSATKD explicitly models and aligns two complementary forms of audio texture: structural textures that capture directional and edge-based patterns and statistical textures that represent distributional characteristics within time–frequency representations. By incorporating these complementary cues during distillation, the proposed framework enables the student model to learn richer acoustic representations that are critical for accurately characterizing complex environmental and underwater sound signals.

Experimental results across multiple datasets demonstrate that combining structural, statistical, and response-based distillation objectives leads to consistent improvements over baseline models and recent knowledge distillation methods. These findings highlight the importance of jointly capturing structural and statistical texture information in acoustic representations, particularly for tasks where fine-grained spectral patterns play a central role in classification performance. Beyond improved accuracy, SSATKD maintains a lightweight student architecture, making it well-suited for deployment in real-time environmental monitoring systems and edge devices where computational resources are limited. Future work will explore extending the framework to additional domains such as bioacoustics and radar as well as integrating self-supervised or multi-modal learning to improve generalization and deployment efficiency.